# Qualitative Research on Software Development: A Longitudinal Case Study Methodology

Laurie McLeod, Stephen G. MacDonell and Bill Doolin

*Auckland University of Technology*
*Private Bag 92006, Auckland 1142, New Zealand*
lauriemcleod@xtra.co.nz, stephen.macdonell@aut.ac.nz, bill.doolin@aut.ac.nz

**Abstract**

*This paper reports the use of a qualitative methodology for conducting longitudinal case study research on software development. We provide a detailed description and explanation of appropriate methods of qualitative data collection and analysis that can be utilized by other researchers in the software engineering field. Our aim is to illustrate the utility of longitudinal case study research, as a complement to existing methodologies for studying software development, so as to enable the community to develop a fuller and richer understanding of this complex, multi-dimensional phenomenon. We discuss the insights gained and lessons learned from applying a longitudinal qualitative approach to an empirical case study of a software development project in a large multi-national organization. We evaluate the methodology used to emphasize its strengths and to address the criticisms traditionally made of qualitative research.*

## 1. INTRODUCTION

Software systems play an increasingly pervasive and important role in contemporary organizations, as well as supporting individuals, and established, temporary and virtual groups. Despite decades of research and the development of an extensive prescriptive literature, as well as some remarkable successes, software systems development projects continue to be problematic, with a significant number failing or performing poorly (Charette 2005; El Emam and Koru 2008; Royal Academy of Engineering 2004; Wallace and Keil 2004). The environment in which software development occurs at present is characterized by a rapid pace of technological change, increased devolution of responsibility and expenditure to user groups, high levels of packaged software acquisition and customization, greater outsourcing of software development, and an increased emphasis on enterprise-wide and inter-organizational software intensive systems (Boehm 2006; McLeod and MacDonell in press). In many cases these changes are interrelated and together demand more flexible, ad hoc or non-traditional development approaches. Combined with the fact that at the same time software systems have become increasingly sophisticated and integrated, the potential for unpredictable or unintended consequences has also increased.

A recent meta-review of prior empirical research on software systems development project outcomes reveals that quantitative factor-based studies form the basis of much of the current knowledge in this area (McLeod and MacDonell in press). Such studies are cross-sectional and attempt to measure the causal influence of a set of contingent conditions or variables on outcomes. However, the continued problematic nature of many software projects and the changing software development environment described above, suggest that there may be a role for alternative research approaches in enabling a deeper understanding of software development processes and practices. A range of research approaches can generate different and richer information about a complex phenomenon such as software development (Harrison et al. 1999; Lethbridge et al. 2005; Mingers 2001). In particular, qualitative research methods have been advocated in this regard, although they still occupy a relatively marginal position in software engineering research compared to quantitative methods (Dittrich et al. 2008; Glass et al. 2002; Segal et al. 2005).

The reason qualitative methods are used relatively infrequently in software engineering research lies in its foundations as a scientific and engineering discipline (Boehm 2006; Harrison et al. 1999). An inherited technical interest in improving practice is associated with a preference for quantitative research approaches that lend themselves to measuring causal relationships for successful software process improvement (Dittrich et al. 2008; Rönkkö et al. 2002). However, studying software engineering is a complex undertaking not just because of its technical aspects, but also "from the awkward intersection of machine and human capabilities, and from the central role of human behavior in software development" (Seaman 1999, p. 557). Accordingly, within the growing trend for empirical studies of software engineering (Perry et al. 2000; Sim et al. 2001; Sjøberg et al. 2007), there are increasing calls for the use of qualitative research methods to address the complex social and behavioral issues and human-technology interactions involved in software development (Dittrich et al. 2007; Lethbridge et al. 2005; Robinson et al. 2007; Runeson and Höst 2009a; Seaman 1999). As the call for papers for this special issue suggests, these issues are central to understanding and improving software development practice.

The main concern in this paper is methodological. Following Robinson et al. (2007), we have two research objectives: (1) to illustrate the utility of one specific approach to qualitative research on software development, longitudinal case study research; and (2) to guide and support other researchers wanting to conduct qualitative research on software development by providing an exemplar of this form of research and a discussion of the lessons learned and issues that arise in conducting such research. To meet these objectives, we describe the qualitative methods used in a longitudinal case study of a recent software development project. The underlying aim of the research was to develop an explanation of how and why the particular project outcomes in the case study emerged over time. A full description and analysis of the case study are beyond the scope of this paper. Full details can be found in McLeod (2008). Instead, aspects of the case study will be drawn on in illustrating the qualitative research methodology followed.

By qualitative research, we mean research that emphasizes words and images rather than numbers and quantitative measurement in data collection and analysis. While many researchers equate qualitative research with an interpretivist epistemology (Bryman and Bell 2007), we apply the term at the level of data or method (Guba 1990). This allows for the possibility of qualitative research within a number of research traditions and with various philosophical assumptions (Dittrich et al. 2007). The main difference in application lies in the research questions asked and in the interpretation of the data. Our own research approach is 'broadly interpretive' (Walsham 1993, 2006), focusing on the content, context and process of software development as a complex and intersubjective social reality that is interpreted rather than discovered. However, essentially the same methods of data collection and analysis that we describe in this paper could be utilized from a post-positivist perspective (Guba 1990). Both traditions seek to conduct research in natural settings close to the phenomena of interest, use qualitative methods to increase richness, and depend extensively on inductive theory grounded in local contexts (Coleman and O'Connor 2007).

The remainder of the paper is structured as follows. In Section 2, we locate the research approach discussed in the paper in the wider context of qualitative case study research. Section 3 describes the selection of the case study used in the paper, a brief summary of the software development project studied is provided, as well as a short comment discussing research ethics, increasingly recognized as an important consideration in software engineering research. Section 4 discusses in detail the methods used for qualitative data collection and qualitative data analysis. In Section 5, we discuss what the longitudinal qualitative research approach enabled us to say about software development. Section 6 evaluates the qualitative research presented in this paper to address the criticisms traditionally made of such approaches. In Section 7, we outline some lessons learned from this research approach that we believe will be useful for other researchers interested in applying qualitative methods to research on software development. Section 8 concludes the paper.

## 2. CASE STUDY RESEARCH

The aim in conducting the research on which this paper draws was to investigate the in situ process and practices of software development as perceived and understood by those involved in performing such work (Coleman and O'Connor 2007; Dittrich et al. 2007). As Lethbridge et al. (2005) argue, this means conducting empirical studies in field settings. That is, the phenomenon under study is investigated "within its real-life context" (Yin 2003, p. 13), using a case study approach. Experiments and surveys are unable to do this to the same extent, since an experiment divorces the phenomenon from its context and surveys are limited in their ability to investigate context (Yin 2003).

Case studies have been a commonly used and legitimate method of research inquiry for studying related fields such as information systems (Benbasat et al. 1987; Cavaye 1996; Doolin 1996; Markus and Lee 1999) and computer supported cooperative work (Dittrich et al. 2007; Wainer and Barsottini 2007) for some time. Equally, the case study method could be usefully deployed in investigations of software development processes and practices, a position that has received recent support (Runeson and Höst 2009a; Segal et al. 2005). After all, software development is generally an (organizational) activity that involves complex interrelationships between people, structure, procedures, politics and culture – elements that require qualitative empirical case study research in real world settings (Runeson and Höst 2009a, 2009b) if they are to be genuinely understood. Indeed, it has been noted that case study research has begun to be used in the field of software engineering, although it often lacks sufficient research grounding or systematic analysis, or is not used to its full potential (Höst and Runeson 2007; Perry et al. 2004; Runeson and Höst 2009a, 2009b; Sjøberg et al. 2007).

The emphasis on studying a phenomenon in its context (Yin 2003) means that case study research focuses on actual organizational tasks, processes and activities, and involves direct contact and close interaction between the researcher and organizational participants (Doolin 1996). Case studies also typically involve multiple data sources, including observation, interviews, documents and archival records, in order to develop a triangulated and in-depth analysis and a contextual understanding of the research setting (Bratthall and Jørgensen 2002; Cavaye 1996; Darke et al. 1998; Easterbrook et al. 2008; Runeson and Höst 2009a; Yin 2003). Case study research offers a degree of flexibility in that key parameters of the research design can be altered during the study in order to react or adapt to "the complex and dynamic characteristics of real world phenomena" (Runeson and Höst 2009a, p. 137).

While case study research does not generate the statistically significant conclusions of controlled empirical studies, it provides rich descriptions of social interactions and practices, reveals the knowledge and perspectives of organizational participants, and yields a "deep understanding of a phenomenon in one context, which may bring insight into others" (Wynekoop & Russo, 1997, p. 51, Easterbrook et al. 2008; Robinson et al. 2007; Runeson and Höst 2009a).

Case studies typically describe a phenomenon, explore a phenomenon, or produce an explanation of a phenomenon (Perry et al. 2005; Runeson and Höst 2009a). *Descriptive* case studies may focus on an extreme or unique case, perhaps a previously inaccessible phenomenon, worth portraying and documenting (Yin 2003). *Exploratory* case studies are used to investigate a little known phenomenon or one without an established theoretical basis. They are often used as an initial exploration to generate new insights or propositions for further research (Easterbrook et al. 2008; Runeson and Höst 2009a). Lethbridge et al. (2005, p. 314) suggest that many case studies in software engineering research are exploratory "because we are still gathering basic knowledge about the human factors surrounding software development and maintenance." *Explanatory* case studies seek to provide a theoretical explanation for a phenomenon. These are most appropriate for addressing "how" and "why" research questions (Yin 2003). Some researchers use explanatory case studies to confirm or refute existing theories (Easterbrook et al. 2008).

In addition, case study research designs can involve a single case or multiple cases. Single case studies may use a critical case to test previously formulated theory or a representative case to inform on typical experiences (Yin 2003). Single case studies may also be used to provide a holistic, in-depth analysis of one setting, characterized by production of the rich descriptions favored by interpretive researchers. In contrast, in a multiple case study design, detail and richness of description may potentially be sacrificed (to some extent) for the opportunity to make comparisons across several settings (Doolin 1996). From a positivist perspective, multiple case studies may represent literal replications (analogous to repeated experiments) or theoretical replications used to generate contrasting results for predictable reasons (Easterbrook et al. 2008; Yin 2003).

### 2.1. Our Approach: Longitudinal Case Study Research

In terms of locating the research approach described and discussed in this paper, a single holistic explanatory case study (Yin 2003) was conducted in order to examine the social and organizational dynamics of software development in situ. The focus was on how software development is enacted as a process, and how and why particular development outcomes emerge over time in particular organizational settings. In order to achieve this consideration of time and context, a longitudinal research design was utilized.

In a longitudinal case study, data are collected over an extended period in order to investigate "how certain conditions change over time" (Yin 2003, p. 42). In-depth, holistic case studies are often carried out longitudinally (Walsham 1993) to facilitate a "multifaceted treatment of change" (Pettigrew 1990, p. 270). A longitudinal case study enables software development to be observed as it unfolds, describing events as they occur and accessing participants' actions and interpretations at the time. This provides rich data that can be analyzed to trace the dynamics of change, which is important when trying to understand the complex interactions surrounding software systems and their development. Software development unfolds within constantly changing contexts and conditions, which are difficult to capture using cross-sectional methods. Longitudinal case studies can support the construction of holistic explanations of the outcomes of complex social processes (Leonard-Barton 1990; Pettigrew 1997; Voss et al. 2002).

In terms of software engineering, longitudinal case studies such as that discussed in this paper address the need for research that focuses on observation of the practice of software developers in order to understand the complexity of software development (Westrup 1993). In doing so, we emphasize the emergent nature of software development rather than the deterministic view prevalent in the software engineering literature, placing "the on-going nature of the software processes … at the heart of the analysis" (Allison and Merali 2007, p. 668).

## 3. CASE SELECTION

In the longitudinal case study described here, the unit of analysis was at the level of a single project. This necessitated identifying and negotiating access to both a host organization and a specific project within it. The contracting between the researchers and a host case study organization is particularly important in longitudinal case study research as there is a need for close, trusted contact over an extended period of time (Pettigrew 1997).

A range of potential research sites was available, as a number of respondents to a survey we conducted into software systems development practice had indicated their willingness to participate further in an in-depth case study. Each of these organizations was examined to establish whether it was an appropriate case study site based on its software systems development profile. This resulted in a short-list of six potential organizations, who were contacted in turn. The first two respondents contacted declined to participate, one because of changed circumstances and the other because of concern over the commercial sensitivity of some of their projects. The third respondent contacted, the CIO of AlphaCo, agreed to meet to discuss the organization's potential involvement in more detail. AlphaCo, a large manufacturing company, operates in a competitive global manufacturing and distribution environment. (AlphaCo is a pseudonym used to refer to the case study organization in this study. Similarly, pseudonyms are used for key individuals, position titles, and organizational units or teams within AlphaCo and other organizations involved in the case study, to preserve confidentiality.)

A series of meetings was conducted with various members of AlphaCo's Information Systems (IS) function, in order to negotiate access to the organization and to identify a suitable project that could be observed as it unfolded. At the initial meeting in mid-December 2004, the CIO was supportive of the proposed case study, suggesting that AlphaCo had a "role in society" to play in supporting research. He agreed to raise the matter with AlphaCo's IS Project Office Manager, who would be able to identify a list of potential projects to observe. After a two month delay, a meeting was finally secured with the IS Project Office Manager, who gave an overview of how software systems projects are executed within AlphaCo and outlined several projects that he thought might be suitable. He agreed to approach the various business owners associated with these projects in order to see who might be receptive to having an external observer on their project. An initial approach to one business unit was declined on the grounds of commercial sensitivity.

After a further month, the IS Project Office Manager successfully identified a potential project and cooperative business owner, and arranged a meeting with a business analyst from the team that manages AlphaCo's information technology (IT) infrastructure outsourcing contract. This team was about to commence a project involving the development of a database solution to replace existing financial spreadsheet models used for managing the IT outsourcing contract. The team member described the project, which seemed to meet the requirements of the research. After some discussion, it was agreed that the first author could observe the project on an ongoing basis as required.

A meeting was then set up with the IS Project Office to negotiate a confidentiality agreement. Part of the difficulty in negotiating an acceptable agreement was that AlphaCo's standard confidentiality agreement was designed for sub-contracted project work and emphasized the ownership of intellectual property arising from the project. It took a series of meetings and emails to negotiate a mutually acceptable understanding that, while preventing the unintentional disclosure of commercially sensitive information was a priority (AlphaCo would be given the opportunity to review the findings of the research to ensure this had not occurred), from an academic research perspective the researchers needed to retain editorial control over the final publication (Pettigrew 1997).

### 3.1. The Project Studied

The project selected for study provided a useful opportunity to investigate software development practices and processes, from multiple stakeholder perspectives, in the modern software systems environment described earlier. It was a short to medium term project, involving commercial software package acquisition, the outsourced development and configuration of that software (including a degree of joint development), a range of internal and external stakeholders and project participants, and devolution of project responsibility to the project business owners. In addition, the project could be observed from initiation to deployment. The approach to longitudinal research taken required a project that could be followed 'in action' (Latour 1987).

The project owners were a small team of business analysts, led by a manager and reporting to a senior commercial manager, who was the project sponsor. Consistent with organizational policy and practice in outsourcing non-core competencies, an external project manager was hired to manage the project. Support for the project within AlphaCo was available from the IS Project Office and an IS Architect seconded to the project. AlphaCo company policy and practice advocated the acquisition of packaged software rather than internal development of a solution. Accordingly, SoftCo, an external development and implementation consultancy, were engaged to build a database solution using a proprietary application development tool (a multi-dimensional database and OLAP tool called MDS), for which SoftCo were the licensed vendors.

The project was intended to be a straightforward migration of existing spreadsheet models to a database solution. From the outset, it was perceived as relatively small and well-defined, supported by senior management, involving committed users and with no major threats to project delivery within the planned six-month project timeframe. In practice, however, the project was subject to various delays and problems so that it stretched over an eighteen month period.

Initial delays were experienced in finding a suitable external project manager, identifying and engaging an appropriate software package vendor, and in negotiating the nature of development. The SoftCo staff member who negotiated the project bid, underestimated the project complexity and solution requirements, committing the SoftCo development team to an extremely tight timeframe and financial budget. The tight project schedule necessitated a lower level of involvement in the solution development than was originally envisaged for the AlphaCo business analyst who would be the primary user of the database solution. Development proceeded in several overlapping and iterative stages, involving cycles of building by the SoftCo developers, testing by members of the AlphaCo project team, and subsequent amendment. Development quickly fell behind schedule, and milestones had to be revised. Even when a largely completed solution was eventually available, final testing was delayed due to data corruption and problems within the solution.

By this time, a major organizational restructuring had taken place within AlphaCo. Not only did the new database solution "fall off the radar", but a change in management focus reduced the demand for reporting from the new solution. Ultimately, the inability to complete the project and produce a usable software solution in a timely manner led to it losing its relevance in the organizational context and its subsequent lack of use within the company.

### 3.2. Ethical Considerations

A number of authors emphasize the importance of ethical standards of conduct for maintaining trust and cooperation in software engineering research (Runeson and Höst 2009a; Sieber 2001; Singer and Vinson 2002). Ethical approval for this research was obtained from the researchers' university ethics committee before fieldwork commenced. During the course of the fieldwork, as people became involved in the project under study or were interviewed, they were spoken to individually by the first author about their potential involvement as research participants. At this point, the person was told about the purpose of the research, the nature of any involvement, what measures would be taken to protect their rights as participants (including protection of their identity), and the right to not participate or to withdraw at any stage. Participants were given an information sheet detailing this information, and were asked to sign a consent form, acknowledging that they understood what was entailed by their participation and agreeing to have various activities audio-taped and used for research purposes. A signed copy of the consent form and information sheet was retained by each participant and by the researchers. These procedures (mandated by the researchers' university) are consistent with those recommended by Runeson & Höst (2009a).

All of the individuals encountered during the case study agreed to participate in the research. Participants engaged with the researcher in a very open, genuine and cooperative way. An overriding concern in conducting fieldwork and subsequent data analysis was to treat their contributions with respect and integrity at all times.

## 4. APPLYING QUALITATIVE METHODS

### 4.1. Data Collection

As noted earlier, case study research involves close involvement with participants, including face-to-face contact, in the research setting. As Nandhakumar & Jones (1997, p. 111) suggest, the best way to understand social processes is "to get access to the actors themselves and to elicit their interpretations directly." In longitudinal case studies, this involves more intensive and extended interaction with participants over a prolonged period of time. This enables observation of social processes as they unfold, provides insight into local knowledge and practices, and reveals alternative or shifting interpretations (Lanzara 1999; Nandhakumar and Jones 1997). A range of data collection methods are available to achieve this familiarity with the organizational context and participants, including observation of activities, conversations and interviews with organizational participants, and review of organizational documentation and project artifacts.

The case study was conducted between mid-2005 and mid-2007. The intention was to follow the project inception, development of the solution, its deployment and initial use. Field work was conducted by the first author. It involved an initial 8-month period coinciding with the main project activity (Figure 1). Work on the project by the various participants was not continuous, so a pragmatic decision was made to observe project work when activities of interest were due to occur. This typically involved the field researcher visiting the site for two or three days each week (averaging about 12 hours per week). However, during a 7-week period of intensive software development activity, she visited the site each working day (averaging about 7.5 hours per day). Subsequent to this initial period of the project, as work on it became more sporadic, a number of follow-up visits to the field site were undertaken until some form of closure could be made in terms of actual use of the new solution. During this second, less intensive, period of the project, regular phone and email contact was maintained with the main participants, in order to keep abreast with progress on the project and to coordinate site visits. In total, 558 hours were spent on site within AlphaCo, observing project activities or interviewing staff.

**Fig 1.** Time spent on site at AlphaCo over the course of the project

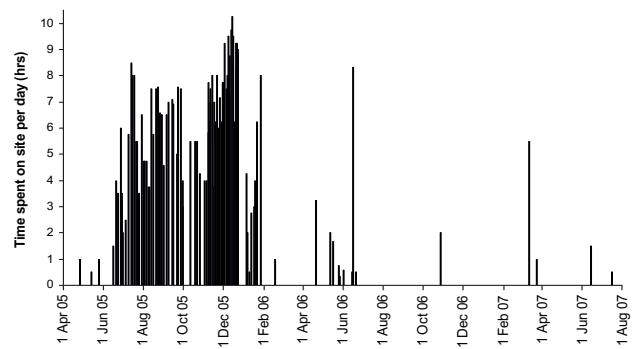

### 4.1.1. Observation

Observation of project activities was a primary source of data in this study. The field researcher (the first author) essentially undertook 'participant observation', in the sense that she participated in the research setting as much as possible over a prolonged period, interacting with the project participants and collecting data with their knowledge (Seaman 1999; Tolich and Davidson 1999). While this did not (usually) comprise direct participation in functional project activities, it did entail close involvement and familiarity with the team at the centre of the study and their daily practices. By personally experiencing the research context under everyday conditions, she was able to acquire sufficient contextual knowledge to be able to interpret what was being observed, facilitating the development of an in-depth 'inside view' on people, issues, events and activities (Bratthall and Jørgensen 2002; Nandhakumar and Jones 1997; Walsham 1995, 2006). From a longitudinal perspective, participant observation over an extended period enabled the observation of software development processes more or less continuously over time.

The field researcher had a desk co-located with the project team for the first eight months of the project, and access to an organizational workspace as required for the rest of the time. She was able to attend meetings, observe project activities, ask questions of and conduct informal conversations with project participants, and access a wide range of documents, including the company intranet and online document repository. The kinds of activities observed included various project management activities based around AlphaCo's established processes, in-house preparation of a solution prototype for use by vendors and the eventual developers, preparation and distribution of a Request for Information (RFI) document, interaction with vendors, product demonstrations, formal selection of a vendor/product, informal and formal project meetings, solution development, testing and training, and solution delivery and transfer to the live environment.

Some 37 formal meetings were held over the course of the project. Observation of these meetings was considered important given their significance as sites of interaction and negotiation between project participants. The field researcher was kept informed of these formal sessions, enabling her to attend all but one. The 36 meetings observed spanned around 47 hours in total (Table 1). Information about the missed meeting was obtained from discussions with participants and reviewing formal documentation.

**Table 1.** Summary of formal meetings observed

| Meeting type | Number observed | Total time (minutes) |
| --- | --- | --- |
| Interviews for an external project manager | 2 | 90 |
| Vendor meetings and demonstrations | 7 | 635 |
| AlphaCo project team meetings | 7 | 450 |
| Weekly project meetings with SoftCo | 9 | 290 |
| Other project meetings with SoftCo | 5 | 175 |
| Training sessions in development tool and solution | 6 | 1175 |
| Total | 36 | 2815 (46.9 hrs) |

As a participant observer, the field researcher and what she was doing were readily accepted by staff at AlphaCo and members of SoftCo. Participants were generally very supportive, often enquiring about the progress of her research or whether she had the support or access she required. Indeed, she was included on the project email list, so as to be kept aware of forthcoming activities or emerging issues. This was particularly important given the discontinuous nature of the site-based field work. Over time, the field researcher developed a close relationship with the project participants and was not excluded from project activities or confidential or sensitive issues. Participant observation included interacting on a social level with members of the project team and staff from SoftCo, such as taking coffee breaks or lunch with them, or being included in conversations and jokes:

> Claire (AlphaCo Business Analyst): Man, this must be boring for you [observing]. It's boring for me, and I'm doing it … Aren't you glad you spent years at university to do that?
>
> Laurie (Researcher): [laughs] Yeah.
>
> (Informal conversation, 14 December, 2005)

A standard joke amongst the project team involved threatening to include her in some of the project work:

> Marie (SoftCo Project Manager): You'll be able to get her [Laurie] to help do some testing. [To Laurie] Frank's got a greater plan for you.
>
> Laurie (Researcher): I don't know the model well enough to do any testing.
>
> Frank (AlphaCo Project Manager): Oh, it's just data validation.
>
> (Informal conversation, 24 November 2005)

From the outset, an open and unconstrained approach to data collection was possible. The primary method of recording field observations was writing detailed notes of what was going on or being discussed, wherever possible using the participants' own words in order to preserve the essence and integrity of what was being said. In total, seven A5 notebooks (each with 200 leaves) were filled with field notes. Field notes were openly written in these notebooks during the observation period as events occurred. At no time was there any need to hide what was being done or to seek "the privacy of the toilets" (Whittle 2005, p. 1308)! In addition, formal meetings, project activities and informal conversations were regularly audio-taped (with permission), particularly where they involved interactions between the AlphaCo project team and the developers from SoftCo. Some 42 hours of recordings were made, providing a comprehensive record of what was said, particularly when more than one participant was involved. Judgment needed to be exercised in deciding which activities to record in order to avoid being overloaded with data. Most audio recordings were transcribed in full; the remainder were listened to and transcribed in part when considered relevant. Participants did not appear to temper what they were saying because the session was being recorded. Comparison of the field notes with transcripts from various audio-taped sessions showed good internal consistency, both in terms of the quality and extent of content that was recorded (Zuboff 1988).

Nevertheless, participants remained aware of the field researcher's observational role and that field notes or recordings were being made, sometimes explicitly referring to them during project activities. For example, during some of the taped sessions, AlphaCo staff would remind the field researcher that certain commercially sensitive material could not be revealed. In another example, in trying to work out how part of the eventual database solution had been created, Claire, one of the AlphaCo Business Analysts, commented, "Your notes will probably say it, somewhere, Laurie" (Informal conversation, 18 January, 2006). Again, in a conversation between AlphaCo and SoftCo project staff about an error in the emerging software solution, a joke was made about the potential use of the researcher's audio recordings:

> Claire (AlphaCo Business Analyst): You had to change a couple of things.
>
> Gary (AlphaCo Business Analyst): The NPV calculation. I don't know what school you went to.

Nancy (SoftCo Senior Developer): Do you know, I wrote that with Frank?

Gary: Did you?

Nancy: Yeah … [To Laurie] You were there.

Laurie (Researcher): I know nothing about NPV.

Marie (SoftCo Project Manager): I can see your little recordings are going to become interesting.

Claire: Very sought after!

(Project meeting, 27 January 2006)

### 4.1.2. Interviews

Another primary source of data was semi-structured interviews, a common form of interviewing in case study research. These involve working from an interview guide – a list of planned questions and interview topics aimed at ensuring their systematic coverage across interviews. However, the interview is flexibly conducted to allow for improvisation and exploration of emerging issues (Allison and Merali 2007; Myers and Newman 2007; Runeson and Höst 2009a).

Very early on, a senior manager said that AlphaCo staff were generally open and that most of them would be prepared to be interviewed as long as the demands on their time were not excessive. This was the field researcher's experience. In order to systematically obtain information relevant to the project, all organizational participants with a direct interest in the software system at the centre of this study and/or those who participated in the project itself were interviewed. Other members of AlphaCo's IS function, chosen because of their knowledge and experience (Zuboff 1988), were also interviewed in order to understand the software systems development environment within the organization. The semi-structured interviews were another important and rich source of data about the project itself, as well as development practice within both AlphaCo and SoftCo.

The interview guides used for the semi-structured interviews were customized to the role of the interviewee, the nature of the information they could provide and the stage the project was at. Interviewees were asked a set of general questions about their organizational role, their involvement or interest in the project, their perspective on organizational events and activities, and their experiences of development practice in their organization. In addition, a list of specific questions was compiled for each participant about aspects of the project that they could comment on, procedural matters, and issues arising during the course of the project. Where relevant, organizational policy and procedure documents or artifacts were used as the focal point of discussion to both elicit information and minimize the potential for misunderstanding.

Some interviewees were interviewed multiple times over the course of the project as new developments or issues emerged. Combining interviews with observation enabled questions to be tailored to the individual experiences of these key informants (Whittle 2005), in an iterative process of observation and verification (Pettigrew 1990).

The longitudinal nature of the research enabled this repeat interview technique, with which participants' changing perspectives and interpretations could be tracked over time (Constantinides and Barrett 2006).

Interviews were typically audio-taped (with permission) and transcribed in full. Where taping was not possible, extensive field notes were made. A total of 33 interviews were conducted, comprising almost 24 hours, for an average interview time of around 45 minutes (Table 2).

**Table 2.** Summary of interviews

| Interviewees | Number of interviewees | Number of interviews | Total interview time (minutes) |
|---|---|---|---|
| *AlphaCo project team* | | | |
| IS Architect | 1 | 1 | 105 |
| External Project Manager | 1 | 5 | 130 |
| Business Analysts | 2 | 12 | 405 |
| Business Managers | 2 | 4 | 135 |
| *Other AlphaCo IS staff* | | | |
| IS Project Office | 3 | 4 | 310 |
| IS Managers | 2 | 2 | 90 |
| IS Analysts | 2 | 2 | 135 |
| *SoftCo project team* | | | |
| Project Manager | 1 | 1 | 30 |
| Senior Developers | 2 | 2 | 90 |
| Total | 16 | 33 | 1430 (23.8 hrs) |

### 4.1.3. Documentary Data Sources

Other data sources were used to supplement the observation and interviewing described above. As noted earlier, full access to project-related emails was ensured by the field researcher's inclusion in the project e-mailing list. These emails were categorized and stored as an additional source of documentary evidence. All project documentation was made available to her through the co-operation of project managers from both AlphaCo and SoftCo. AlphaCo's IS policy, planning and process documents were also reviewed so that the field researcher could become familiar and fluent with IS processes, practices and terminology. Much of this documentation was contained in a digital process document repository. A range of organizational documents and electronic sources provided a rich source of contextual information. Examples of the organizational and project documentation reviewed are shown in Table 3. In addition, over 100 publicly available articles on AlphaCo and its IS function were accessed and reviewed. The digital nature of almost all of the documentation reviewed facilitated its management and manipulation in the subsequent qualitative data analysis.

### 4.2. Data Analysis

Preliminary data analysis began early in the project and continued during the main period of intensive field work and after as required. It entailed the field researcher repeatedly reviewing the field notes and other documentation available at the time in order to identify relevant activities, events and issues related to the project. These were recorded in a spreadsheet together with the sources of data that informed them, including which participants' views had been obtained or still needed to be sought. The spreadsheet served as an important resource drawn upon by the field researcher to manage data collection from project participants and other data sources.

**Table 3.** Organizational and project documentation reviewed

| AlphaCo organizational documentation | |
|---|---|
| Organizational project management standards | Staff intranet |
| Organizational restructuring documents and presentations | Internal company monthly magazine |
| Organizational website | Company annual reports |

| AlphaCo IS process documentation | |
|---|---|
| IS process document repository (procedures, templates, checklists, guidelines, examples) | IS strategic and business plans |
| | IS quality standards and governance documentation |
| IS project life cycle | IS balanced scorecard reports and documentation |
| Project tracking system and manual | |
| IS policy framework and policies | |
| IS induction manuals | |

| AlphaCo project documentation | |
|---|---|
| IT outsourcing contract management documentation | Vendor reference checks |
| | Vendor evaluation reports |
| IT outsourcing contract management financial models and documentation | Vendor service level agreement |
| External project manager Terms of Reference | Project deliverables (concept document, project definition, feasibility reports, project plan, closure report) |
| Solution prototype, data and documentation | Project monthly progress and update reports |
| RFI document | Project self-assessment report |
| Vendor RFI responses and product information | |

| SoftCo project documentation | |
|---|---|
| Project costing and deliverables documentation | Training manuals |
| | Solution documentation |
| Project meeting agendas and minutes | Software license agreement |
| Task allocation plan | |
| Issues register | |

After the end of the main data collection period, a more comprehensive, thematic analysis was used to interpret the data collected. Thematic analysis is a widely used method for identifying, organizing and reporting patterns or themes that capture meaningful aspects of the data in relation to a research question (Braun and Clarke 2006). Themes may be generated inductively (from the data) or deductively (from a priori theory). In the study described here, a combination of these approaches was used. The data from the various sources described above (i.e. field notes, interview transcripts, emails, and various project and organizational documents) were progressively collated into electronic form in documents. Initially, data were broadly categorized using eighteen factors influencing software systems development that had been identified from prior literature (McLeod and MacDonell in press). For ease of management, extracts from the data relating to each initial category were contained in a separate document. Data extracts were identified by reference to their source (e.g. collection date and page number in a field notebook, or the source electronic document). Where a data extract was considered to be relevant to more than one category, it was included in multiple locations (Whittle 2005). Cross-references were made between different categories to reflect and reinforce their inter-relatedness and interaction.

Data within each initial category were read and re-read on multiple occasions, often separated by significant periods of time. In this part of the analysis, the data were categorized and re-categorized using a process of constant comparison (Glaser and Strauss 1967) to identify more refined themes based around specific aspects of how the project was enacted. These themes were predominantly inductively derived from the data, although the analysis was informed by the researchers' understanding of software development and by reference to relevant social theory and the findings of prior empirical studies. As a more detailed understanding of and familiarity with the data was achieved, progressively more detailed themes (often based around the research participants' own vocabulary) were applied to the data and used to organize it. In this way, the data analysis was an emergent process involving interplay between data interpretation and theory (Walsham 1995, 2006). Table 4 illustrates the range and depth of themes that could develop around a given initial category, in this case project participants' 'understanding' in interaction around software development.

Being able to manipulate data electronically was a convenient way of handling, searching and comparing the large volumes of data involved. Co-locating extracts of data from various sources in their entirety (as opposed to single codes) in the same document meant that the researcher revisited the textual context in which the data was situated each time the data was read. The aim of this was to minimize the possibility of making interpretations based on the data removed from the context in which they occurred. Throughout the data analysis, explicit attempts were made to retain chronological order and temporal relativity of the data, in order to facilitate the subsequent identification and description of episodes within the case study. One technique used was to visually represent across elapsed project time critical events and changes in aspects of the project content, context, participants and their interactions, and software development processes. These visual representations were an important resource in interpreting the case study data and structuring the subsequent case description and narrative.

Data from the thematic analysis formed the basis of a detailed 'chronological' (Allison and Merali 2007; Yin 2003) case narrative of the project as it unfolded over time. Twenty-one key project activities were identified, together with critical events and changes in aspects of the project content, context and process Pettigrew 1990; Walsham 1993) over the course of the project. Temporal bracketing (Langley 1999) was used to divide the case narrative into eight episodes, each reflecting a degree of continuity of activities in the software development process and distinguishable from other episodes for analytical purposes. The use of episodes both structured the narrative and allowed the researchers to consider actions and interactions in one period in the context of the larger temporal whole or interconnectedness (Pettigrew, 1990).

**Table 4.** Themes developed in this project around the initial category 'understanding'

| | Theme | | | | | Theme | | |
|---|---|---|---|---|---|---|---|---|
| − | **Conceptions of the project** | | | | − | **Understanding the original spreadsheet models** | | |
| | − | As a database solution | | | | + | For all | |
| | | + | Within AlphaCo | | | + | For Claire | |
| | | + | Within SoftCo | | | + | For Frank | |
| | | + | Within Vendor1 | | | − | For Gary | |
| | + | As a tool | | | | | − | His understanding of the models |
| | + | Building/constructing a solution | | | | | | − Of the evaluation model |
| | − | Configuration | | | | | | − Of the scorecard model |
| | − | Translation | | | | | − | Source of that understanding |
| | + | Migration of existing model | | | | | | − Through his involvement in the project |
| | + | Replication of existing model | | | | | | − Asking questions |
| − | **Understanding the process** | | | | | | | − Using the model |
| − | **Barriers to understanding** | | | | | | − | Consequences for the project |
| | − | Time/Tight deadline | | | | | | − From his understanding of evaluation model |
| | − | The model itself | | | | | |   − Using the model in MDS |
| | − | Access to Claire's original model | | | | | |   − Contributing to others' understanding |
| | | − | Concerns over not accessing C's model | | | | |   − Role in testing |
| | | − | Were SoftCo's concerns justified? | | | | |   − Contributing to form of emerging model |
| | − | Not having all the information at the outset | | | | | |   − Caused things to drag at the end |
| | − | Things not 100% defined | | | | | | − From his understanding of scorecard model |
| | − | Coming in late and under-prepared | | | | | − | Consequences for ongoing use of MDS model |
| + | **Aspects of the model causing difficulty** | | | | | | | − From his understanding of evaluation model |
| | + | Understanding scenario copying | | | | | | − From his understanding of scorecard model |
| | − | Resource calculations | | | | − | For SoftCo | |
| | | + | These are complex | | | | − | Understanding of the models |
| | | + | Getting it right | | | | | − | Consequences of late/partial understanding |
| | | + | Getting it right for Year 2 onwards | | | | | | − Inefficiencies |
| + | **Understanding the MDS technology** | | | | | | | − But model largely the same |
| − | **Understanding the emerging model** | | | | | − | Source of that understanding | |
| | − | The overall model | | | | | − | What they did not do |
| | − | Developing the emerging model | | | | | | − Project definition |
| | + | Aspects of the model itself & how it is built | | | | | | − Missing the scoping meeting |
| + | **Misunderstanding** | | | | | − | What they did do instead | |
| | | | | | | | − Using the solution prototype | |
| | | | | | | | − Diagrams from Claire | |
| | | | | | | | − Interactions | |
| | | | | | | + | The process of creating understanding | |

+ Indicates further detail or sub-themes
− Indicates no further detail under the theme or sub-theme than that shown

Rather than simply a case history or description, the case narrative attempted to provide a meaningful explanation of how and why the sequence of events in this particular software development project, situated in a specific organizational setting and involving multiple groups of internal and external actors, produced the observed outcome (Pentland 1999; Pettigrew 1997). The intention was "to derive theoretical interpretation from data … rather than to test theory against data" (Nandhakumar and Avison 1999, p. 180). Central to this theoretical interpretation was a conceptualization of software development as a complex and emergent process, often with unintended effects, involving a dynamic relationship between people, software technology, and the social and organizational context in which development occurs (Allison and Merali 2007; Constantinides and Barrett 2006; Gasson 1999). Participant quotations were used in the case narrative to illustrate important points in the analysis. Typically, these were selected as representative of the views of the participants involved. Where appropriate, unique or contrasting views were also presented (Zuboff 1988).

To support and summarize the case narrative, 'visual mapping' (Langley 1999) was used to produce a diagrammatic representation of the sequential relationship between episodes over the course of the software development project. This 'process chart' (Langley and Truax 1994) depicted at a high level the emergence of the project outcome over time from a trajectory of key (often overlapping and iterative) project activities, influenced by significant contextual events, project participants' perceptions and actions, and material or technological artifacts. As such, the process chart provided a useful abstraction and temporal representation of the case study data that could be used to support the more detailed case narrative. Figure 2 shows one section of the process chart from the study to illustrate the technique (for the complete process chart and a fuller description of the approach used, see McLeod (2008)). It includes three of the analytically identified episodes, which relate to the identification of SoftCo as a potential vendor, negotiation of how development would proceed, and the initial development of a solution. The six key project activities included in these episodes are briefly summarized in Table 5. The graphical notation used in the process chart is adapted from Langley and Truax (1994) and Madsen et al. (2006).

**Fig 2.** Partial process chart for the AlphaCo software development project

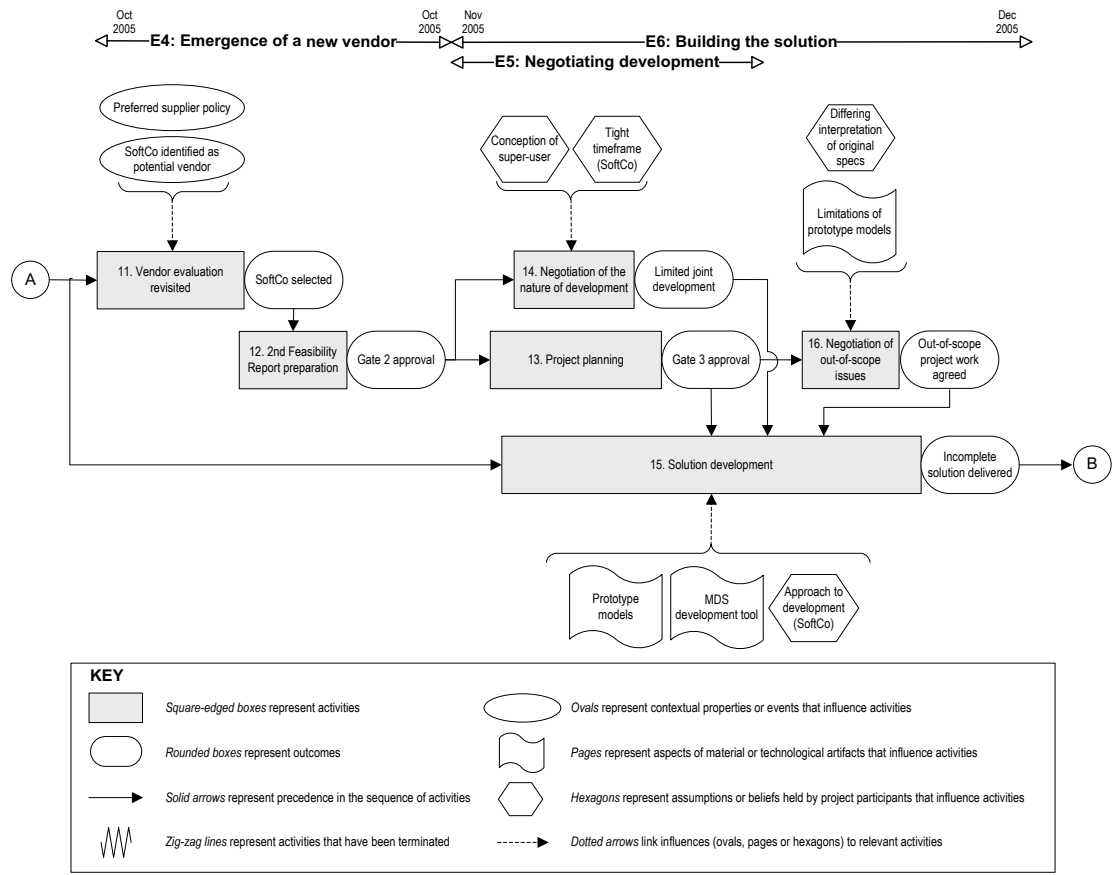

**Table 5.** Key project activities in Episodes 4-6

| Episode 4 | **Emergence of a new vendor** |
|---|---|
| Activity 11 | *Vendor evaluation revisited*: The inadvertent discovery of SoftCo, with a product (MDS) already used by an AlphaCo subsidiary, led to a re-evaluation of the available solution options. MDS was constructed as the preferred solution on cost since software licenses and a hardware server could be shared with the subsidiary. |
| Activity 12 | $2^{nd}$ *Feasibility Report preparation*: This project deliverable was rewritten and approval to proceed was granted. |
| **Episode 5** | **Negotiating development** |
| Activity 13 | *Project planning*: Initial planning with SoftCo established that development was to be done onsite at AlphaCo on a standalone server with considerable input from members of the AlphaCo project team to transfer knowledge of the MDS tool. A development timeframe was agreed. To save time, agreement was obtained from the IS Project Office to start development without waiting for formal approval of a detailed project plan. |
| Activity 14 | *Negotiation of the nature of development*: The extent of the AlphaCo project team members' participation in the solution build was renegotiated and reduced due to SoftCo concerns about tight project costs and timeframe. |
| **Episode 6** | **Building the solution** |
| Activity 15 | *Solution development*: Development proceeded in overlapping and iterative cycles of building, testing and amendment. Development quickly fell behind schedule, and milestones were revised. |
| Activity 16 | *Negotiation of out-of-scope issues*: A SoftCo perception of escalating project scope and costs led to the negotiation of whether remaining project tasks were part of the original project specifications or 'out of scope'. |

# 5. WHAT DID THE LONGITUDINAL QUALITATIVE APPROACH YIELD?

We believe that using a longitudinal qualitative case study methodology in software development research opens up possibilities that have not received extensive attention to date in the software engineering field. In this section, we outline some of these possibilities in terms of what this research approach yielded in the case study on which the paper is based. We identify three main contributions: the production of holistic explanations of software development as an emergent process, the derivation of specific insights about software development practice, and the generation of theoretical models of software development.

## 5.1. Software Development as a Complex and Emergent Process

The invariant relationships between variables and outcomes assumed by traditional factor studies rarely exist in complex, 'real world', social phenomena. Reductionist approaches to dealing with software development fail to adequately deal with the dynamic and interactive processes involved in such a complex organizational practice. In contrast, a longitudinal qualitative approach such as that presented here opens up the 'black box' (Latour 1987) of software development, by focusing attention on the sequence of events and situated actions that comprise the software development process and that connect antecedents with outcomes (Pentland 1999), "enabling one to track cause and effect" (Leonard-Barton 1990, p. 250). Software development trajectories are rarely linear, continuous and predictable. Instead of singular explanations of software development, longitudinal case study research facilitates the production of holistic and analytically complex explanations that take into account the discontinuous, multi-directional and open-ended nature of the development process as it emerges over time (Langley 1999; Pettigrew 1997).

In the case study described here, the software development project did not take the smooth and straightforward path anticipated by its originators. Rather than following the intended linear project lifecycle, the project trajectory was characterized by overlapping and parallel tracks of events and activities, reverses or iterations in specific aspects of development, and was subject to unintended consequences or unanticipated events. Instead of a predictable project path and outcome, the software development that occurred in the case study was actually anything but that, experiencing delays and difficulties more typical of larger projects. The project's trajectory and outcome could not be adequately explained in terms of simple prescriptions for successful development or the presence or absence of particular factors. The protracted and problematic nature of development that led to a software solution that had lost its organizational relevance was an emergent outcome or effect of complicated, multi-dimensional interrelationships and interactions between social, material and contextual influences on software development.

## 5.2. Insights into Software Development Practice

The qualitative case study analysis provided a number of insights into the nature of software development that we believe are of potential interest to software development in other locales. These insights arise from the richness of the data obtained, the detailed and in-depth analysis this enabled, the longitudinal dimension incorporated into the study, and the consideration of the content, context and process of software development. While the insights generated in the case study analysis are not necessarily novel, they contribute towards building a body of cumulative research that addresses the complex and dynamic nature of software development, which can better inform software development practice and lead to more beneficial outcomes for all stakeholders involved.

### 5.2.1. The Influence of Initial Characterizations of a Project

Initial characterizations of a software development project can play an important role in shaping participants' attitudes and actions in the development process. The early characterization of the project in this case study as small, well-defined and uncomplicated meant that decisions and choices were made that affected the project trajectory. On AlphaCo's part, no baseline review of the adequacy of the original financial models was undertaken, the decision was made to forego a formal problem definition process, and the employment of an inexperienced external project manager was considered acceptable. A similar underestimation of the complexity of the project by SoftCo committed them to an overly tight timeframe and project budget, with consequences for how the nature of development was negotiated and indeed how the solution development proceeded.

### 5.2.2. The Role of Participants' Sense-Making in Shaping Actions and Decisions

Consideration of how individual project participants' knowledge, expectations, perceptions and interests shape their sense-making, decisions and actions helps to explain how software development proceeds through the negotiation and communication of a shared understanding of the development problem and solution. In the case study, this often involved a particular interpretation of the problem at hand that stabilized its meaning and aligned the understanding and interests of different participants around it (Gasson 1999). Ambiguous or problematic aspects of the software development process were often conceptualized and made sense of by the use of various symbolic artifacts such as metaphors or 'transient constructs' (Lanzara 1999), enabling a way forward to be identified and negotiated by the participants. For example, the construction of the notion of 'out-of-scope' project work by the SoftCo project manager provided a way of understanding and making sense of contested elements of the project. It enabled the negotiation of what tasks were legitimate demands on the developers' time, how project costs would be allocated, and what was an acceptable solution delivery timeframe. The case study analysis also highlighted how the contribution or lack of contribution of key individuals can facilitate or constrain the development of a shared understanding in software development.

### 5.2.3. The Material Content of a Project

The case study analysis highlighted the importance of not neglecting the technological dimension in analyses of software development. Material resources and technological artifacts mediate the work practices and interactions of the various participants involved in the software development process (Schultze and Orlikowski 2004). At a general level, the material and technological content of the project studied, such as the chosen development technology and existing IT infrastructure, functioned as constraints and capabilities within which the emerging software solution was developed. In addition, particular project-related artifacts and representations (e.g. the solution prototype, RFI document, task allocation plan, issues register) functioned as 'boundary objects' (Star 1989; Subrahmanian et al. 2003) in facilitating communication and collaboration between the AlphaCo project team and the SoftCo developers. Boundary objects span different communities or groups, providing sufficient flexibility in interpretation to accommodate individual meanings and interests while serving as a common basis for knowledge-sharing and negotiation. In the case study, particular boundary objects stood in for or delegated for individual participants and their knowledge. Other boundary objects were used as the basis for contract negotiation or conflict resolution. From a practice perspective, software engineers need to understand the potential for such objects to influence interaction and negotiation between participants in a software development project.

### 5.2.4. The Importance of Context

Consideration of the context in which a software development project is situated helps to demonstrate how various contextual elements influence and structure project-related activities by providing rules and resources that participants draw on in making choices and decisions, thus shaping the field of possible actions (Giddens 1984). The case study analysis highlighted how, for example, AlphaCo's guiding principles for software systems acquisition and development, standard procedures for project management, and historical organizational practices were implicated in shaping the software development process and trajectory in particular ways. Further, software development occurred in an organizational context of rapid and continuous change. The longitudinal analysis revealed how the 'formative context' (Ciborra and Lanzara 1994) within which the proposed software solution was situated shifted over time, thus undermining the relevance of the developed solution within a changed organizational context.

### 5.2.5. The Role of Unanticipated Events and Unintended Consequences

Finally, the case study analysis reveals the extent to which the software development trajectory and emergent solution may be shaped and influenced by unanticipated events or unintended consequences of decisions and actions (Markus and Robey 1988). For example, the inadvertent discovery of SoftCo, late into the vendor engagement process, delayed the project and introduced a new set of actors with various associated consequences. In another example, a decision to share software and hardware with an AlphaCo subsidiary in order to reduce costs led to extensive time delays and problems in transferring the eventual software solution from the development environment to AlphaCo's networked environment. Multiple or ongoing unintended effects, often of varying nature, can arise from a single decision or action. One example from the case study was the underestimation of the project complexity by SoftCo mentioned above. This committed SoftCo to unrealistic project milestones and costs, with consequential downstream effects in the form of project slippage, version control problems, solution errors, inadequate testing and repeated amendments. It also meant development was renegotiated to reduce AlphaCo team members' participation in development work, which compounded the delays experienced and reduced the potential for transfer of knowledge of the MDS development tool.

Software development tends to be portrayed as a predictable and controllable intervention in the software engineering literature (Allison and Merali 2007). In contrast, the case study analysis supports the contention that control or management of a software development project is an emergent property of situated software development practice, rather than a determining factor (Madsen et al. 2006), as participants recognize and respond to unanticipated events or the unintended consequences of decisions taken (Galliers and Swan 2000).

### 5.3. Towards a Theoretical Model

In analyzing the case study, it became apparent to the researchers that the project trajectory and outcome could not readily be reduced to a single set of contributory factors. Rather, a more detailed consideration and conceptualization of the complex practices and processes involved in software development was required. Overall, the analysis enabled the development of an understanding of software development as a process in which a software system emerges from a dynamic and interactive relationship between the technological and material components of the software system and artifacts used in development, the social and organizational context in which the software system is designed, developed and used, and the negotiated actions of various individuals and groups with an interest in the software system or its development.

Based on this understanding, a theoretical model of software development as situated practice was developed

(see McLeod 2008). This was a major outcome of the research, and was used to analyze, interpret and present key events and activities in the case study analysis. It provided the theoretical basis for the explanation of the software development process outcomes generated in the study. It is also of potential value outside the study, as an analytical tool for other software engineering researchers, as a guide for project managers in planning and executing software development projects, as a basis for software project evaluation, or in educating future software engineers about the situated and emergent nature of the software development process in practice.

## 6. EVALUATING THE RESEARCH

Qualitative research and case studies are often criticized as being of limited value (e.g. useful only for exploratory research), subject to bias, and inadequate as a basis for generalization (Flyvbjerg 2006). Runeson and Höst (2009a) suggest that this critique can be addressed in two ways. The first is to ensure that appropriate research methods and practices are systematically and consistently applied. Runeson and Höst emphasize the importance of adequately defining and describing the data collection and analysis methods used, providing a chain of evidence with traceable inferences, and the consideration of alternative perspectives and explanations. They note that this involves more than simply listing methods; rather a detailed 'history of the inquiry' is called for. The description of the application of qualitative data collection and analysis methods in the study provided earlier in the paper is such an account.

In addition, the validity of qualitative research can be improved by using triangulation, as multiple sources of data are likely to result in better justified findings (as well as more findings) (Bratthall and Jørgensen 2002; Dittrich et al. 2007; Easterbrook et al. 2008; Runeson and Höst 2009a). As Miller (2008, p. 226) observes, "Triangulation provides an explicit vehicle for tackling the principle issues or limitations presented by a single empirical study." In the study described here, multiple and complementary sources of data were used and multiple perspectives across different levels in the organization were sought. Validity was also improved by spending sufficient time with the case, and analyzing negative or contradictory evidence (Easterbrook et al. 2008; Runeson and Höst 2009a).

The second way that the traditional critique of qualitative research and case studies can be addressed is by understanding that statistical significance is not the only form of valid knowledge (Runeson and Höst 2009a). For example, Yin (2003) argues that analytic, not statistical, generalization is the goal of case study research; case studies "are generalizable to theoretical propositions and not to populations or universes" (p. 10). Walsham (1995) extends this in arguing that such research can be used to develop theoretical concepts that inform further theoretical development, to generate or refine theoretical frameworks, to draw specific implications from one particular domain that can be useful in understanding similar phenomena in other contexts, and to contribute rich insights or implications on a wide range of issues. The longitudinal case study of software development described here developed a number of theoretical insights and an in-depth understanding of software development as an emergent process that, although grounded in the particularity of the case study, are nevertheless relevant and meaningful beyond the research site.

Research should be judged according to criteria appropriate for the underlying assumptions of the research tradition within which it is conducted, and there is growing recognition that interpretive case studies, such as the research described in this paper, should be evaluated against criteria appropriate to their nature (Klein and Myers 1999; Markus and Lee 1999; Walsham 2006). The validity of the understanding or interpretation derived from such research relies on its ability to provide a convincing explanation of the phenomena studied and on the clarity of the logical reasoning underpinning its argument (Walsham 1993). Following Golden-Biddle & Locke (1993), Walsham & Sahay (1999) suggest that to produce convincing explanations, interpretive case study accounts need to demonstrate authenticity, plausibility, and criticality. The way in which these three criteria were addressed in analyzing and presenting the longitudinal case study is summarized in Table 6.

**Table 6.** Criteria for evaluating the longitudinal case study (after Walsham and Sahay 1999)

| Criterion | Explanation | Application in the study (McLeod 2008) |
|---|---|---|
| Authenticity | Shows that the authors have 'been there', by conveying the vitality of life in the field | • Describes the extent and nature of fieldwork, the role of the researcher and interaction with participants<br>• Rich descriptions that display familiarity with participants' everyday actions and use quotes from participants<br>• Seeks multiple participant perspectives and is sensitive to different participant interpretations<br>• Demonstrates systematic, disciplined and iterative approach to data collection and analysis |
| Plausibility | Connects to the personal and professional experience of the reader | • Uses schematics such as tables, figures, models and visual mapping to make sense of the data for the reader<br>• Develops a comprehensive explanation of the phenomenon studied that is supported by relevant theory<br>• Explicitly considers the social and historical context of the phenomenon investigated<br>• Demonstrates the relevance of the analysis to contemporary software development practice in organizations<br>• Makes a distinctive contribution through the development of an analytical model of software development |
| Criticality | Encourages readers to consider their taken-for-granted ideas and beliefs | • Offers a novel model for understanding software development as situated practice<br>• Applies non-mainstream ways of thinking about social interaction and the role of technology in software development. |

# 7. LESSONS LEARNED

Dittrich et al. (2007, p. 536) suggest that an important step in qualitative research involves "reflection on the lessons learned and the applicability and usefulness of the chosen research design". In this section, we reflect on the longitudinal qualitative approach used in the case study and presented in the paper to derive a number of lessons learned that we hope will be helpful to other software engineering researchers and in improving the relevance of qualitative research on software development (Allison and Merali 2007).

## 7.1. Researching in Action

The key benefit of the longitudinal qualitative approach taken in the study was the ability to follow software development in action, before the software system is completed and "blackboxed" (Latour 1987, p. 258). Observing software development processes continuously over time is essential to producing a temporal understanding and explanation of how processes are connected to outcomes in an emergent and often unpredictable way. While a longitudinal dimension can be added retrospectively, such studies rely on the recall of participants, which may be faulty or susceptible to influence from subsequent events or post-rationalization (Leonard-Barton 1990). Further, researching in action reveals the way that participants' perspectives, interpretations or positions may change over time.

Despite the benefits, there are practical considerations associated with this type of research. Observing processes continuously over time requires considerable access to and time spent at a research site (Nandhakumar and Jones 1997; Westrup 1993). For various reasons, it may be impractical for a researcher to be in the field for the entire period of software development activity. Choosing when and what to observe may be somewhat arbitrary, unless the research participants can assist in keeping the researcher informed – more likely if a close and cooperative relationship with participants is established. Even then, there is the possibility of unanticipated events occurring when the researcher is not present, necessitating reliance on retrospective participant accounts. However, as experienced in this case study, the intensity of activity in software development projects can vary over time, which may mean that fieldwork can also be conducted in stages of varying intensity to match development project activity.

The access involved in a longitudinal qualitative study requires a considerable commitment from the participating organization to cooperation over an extended period of time. To the host organization, the costs of the research in terms of time, disruption and commercial confidentiality may seem to outweigh the perceived benefits, necessitating much work by the researcher in managing organizational expectations and relationships (Leonard-Barton 1990; Nandhakumar and Jones 1997). Even if access to a suitable organization is secured, access to potential projects may not be under the control of your organizational sponsor, or commercial sensitivity may preclude outsiders accessing particular projects or parts of an organization. Further, conduct of the research can be adversely affected by events within the organization. In the case study described here, time and organizational restructuring meant that many of the key participants left the project (or indeed the organization) before fieldwork was eventually completed.

## 7.2. Rich Data

Close engagement in a particular research context yields the rich data needed for an in-depth and holistic analysis based upon the 'thick description' (Geertz 1973) of software development processes (Allison and Merali 2007; Nandhakumar and Jones 1997). As noted earlier, prolonged and close proximity to the phenomenon being researched and the use of multiple methods, including participant observation, semi-structured interviews and document review, produced a familiarity with and understanding of the content, context and process of software development in the case study. Participant observation meant that useful data was obtained from informal conversations and observation during software development activity as well as from the semi-structured interviews. The former were particularly important for deriving a context-sensitive understanding of the project studied and also for discerning the sometimes subtle differences between individuals' perspectives and interests within the respective project teams.

However, as various authors have noted, the collection of large amounts of rich data in longitudinal qualitative research can be as challenging as it is potentially rewarding. Such large volumes of complex data require time and effort to analyze. It is easy to be overwhelmed by the amount and richness of the data, or so focused on detail that higher-level interpretations or patterns do not emerge (Leonard-Barton 1990; Nandhakumar and Jones 1997; Robinson et al. 2007). Still, as Seaman (1999, p. 558) points out, if qualitative data are "more difficult to summarize or simplify … then, so are the problems we study in software engineering." In the case study described in this paper, a number of mechanisms were used to deal with the volume and complexity of data generated from the fieldwork. These included the use of a spreadsheet by the field researcher to manage data collection around specific project activities, events and emerging issues across the various data sources available. Temporal bracketing was used to divide the project trajectory into analytically meaningful episodes of related software development activities, and key elements of the process data were abstracted out and graphically represented using visual mapping techniques.

## 7.3. Multiple Perspectives

Actively accessing multiple participant perspectives is important in developing a holistic understanding of an organizational phenomenon like software development, particularly as it often involves a wide range of internal and external participants and stakeholders. Multiple perspectives across different organizational levels are important in obtaining data relevant to multiple levels of context, and in avoiding reliance on or bias from a single perspective. In the case study described in this paper, the AlphaCo managers were able to provide information about, for example, organization-wide policies and issues to do with IT strategy or software acquisition and development. This 'big picture' perspective (Coleman and O'Connor 2007), although useful, was removed from actual software development work. To understand the organizational history of software development and how it occurred in practice within AlphaCo (as opposed to policy), it was necessary to talk with those participants for whom software development was part of their everyday experience and work. Similarly, conversations with the SoftCo developers revealed a much more informal knowledge of and adherence to the software development methodology espoused by the organization and the SoftCo project manager. Again, discussing the software development project with the externally-appointed AlphaCo project manager provided a refreshing and often counter view of organizational and software development practice in AlphaCo.

Sensitivity to multiple participant perspectives applies not just across different hierarchic levels in an organization or different groups associated with software development, but also within groups. Commonly used categories such as 'users', 'developers' or 'managers' are not homogeneous groups, but comprise individuals with potentially different characteristics, interests and capabilities. Further, individuals can have multiple or changing roles in a software development project and their actions may be influenced by a range of personal, organizational or professional commitments and affiliations (McLeod and MacDonell in press). For example, the field researcher observed how, during the project in this case study, a relatively junior AlphaCo business analyst, who had been given an official status as a 'super-user' of the MDS development tool being acquired from SoftCo, successfully resisted this technical definition of his role in favour of a continuation of his current business role.

Multiple participant perspectives may also imply differences, or even conflicts, in participants' interpretations of events and activities. Rather than attempting to resolve such differences, perhaps based on the perceived reliability of various participants, in order to arrive at a singular explanation, the approach taken in this study was to exploit and tease out differences and their reasons to reveal differing aspects and a more holistic explanation of the software development process studied. For example, in this case study, tension arose between the AlphaCo and SoftCo project teams around whether particular development tasks were in or out of scope. Observation of this tension and subsequent conversations with the participants revealed different interpretations of not only the project specifications but also the basis for how development should have proceeded. This difference in understanding was crucial in explaining why the software development trajectory proceeded as it did in the case study.

## 7.4. Relationship between the Researcher and the Researched

A number of authors have highlighted the significance of the relationship between the researcher and the researched in qualitative research involving close engagement with the research context (Nandhakumar and Jones 1997; Robinson et al. 2007). These authors suggest that there is a "tension involved in moving between two worlds" (Robinson et al. 2007, p. 545), arising from the dual role of the field researcher as both a participant experiencing the research context and a researcher analyzing and interpreting it. This had several consequences in this study, including negotiating and maintaining an appropriate distance (or closeness) to the research context and participants. For example, the field researcher was comfortable socializing with participants at the workplace or during work hours, but turned down an invitation to an after-hours Christmas party organized by SoftCo for their clients. Similarly, she was careful to maintain an association with all participants, to avoid being perceived as favoring specific individuals. Care was also taken not to disclose to others personal information or views expressed to the researcher by any one participant.

In many ways, being known to be a doctoral researcher was advantageous to conducting field work in the software development project. It was a role the participants recognized and seemed to respect, and helped the researcher to engage the participants' interest and cooperation while maintaining a certain degree of distance or difference. That AlphaCo staff were used to having external parties involved in software development projects may also have facilitated acceptance of the field researcher and her participation in the organizational context.

## 8. CONCLUSION

In this paper, we outline the longitudinal qualitative research approach used to investigate a case study of a software development project in a large manufacturing organization. We discuss the utility of this approach and show how it enabled the derivation of an in-depth understanding of software development as a complex and emergent process, the generation of a range of insights into software development practice, and the development of a theoretical model of software development as situated practice that was used to explain the outcome of software development in the case study but is also of potential value in other contexts. We also abstract out some lessons learned in conducting the research that may benefit other qualitative researchers in software engineering.

In presenting in some detail the qualitative research methods that were followed in the study, we provide practical guidance on one way to undertake qualitative research on software development practice. Researchers can use the approach developed here to conduct, analyze and illustrate the processes occurring in the longitudinal case studies of software development. Comparison of how software project outcomes emerge and unfold over time in multiple case studies will help "build up a repertoire of knowledge about what can be expected in practice and what can be done to cope with the situation" (Madsen et al., 2006, p. 236).

The analysis summarized in the paper confirms the utility of qualitative approaches for addressing the complexity of software development processes as enacted in local settings. A major implication of this study is that a multi-dimensional consideration of software development is needed to understand this complex organizational phenomenon. Adopting such a research approach facilitates a holistic analysis rather than a narrow focus on individual dimensions and aspects of the phenomenon.

# ACKNOWLEDGMENTS

This research was funded through a Top Achiever Doctoral Scholarship by the Tertiary Education Commission of New Zealand. We would like to acknowledge the support of AlphaCo and SoftCo.

# REFERENCES

Allison I, Merali Y (2007) Software process improvement as emergent change: A structurational analysis. Inf Softw Technol 49(6):668-681

Benbasat I, Goldstein DK, Mead M (1987) The case research strategy in studies of information systems. Manag Inf Syst Q 11(3):369-386

Boehm B (2006) A view of 20th and 21st century software engineering. In Rombach D, Soffa ML (eds) Proceedings of the 28th International Conference on Software Engineering. ACM, New York, pp. 12-29

Bratthall L, Jørgensen M (2002) Can you trust a single data source exploratory software engineering case study? Empir Softw Eng 7(1):9-26

Braun V, Clarke V (2006) Using thematic analysis in psychology. Qual Res Psychol 3(2):77-101

Bryman A, Bell E (2007) Business research methods, 2nd edn. Oxford University Press, Oxford

Cavaye ALM (1996) Case study research: A multi-faceted research approach for IS. Inf Syst J 6(3):227-242

Charette RN (2005) Why software fails. IEEE Spectr 42(9):42-49

Ciborra CU, Lanzara GF (1994) Formative contexts and information technology: Understanding the dynamics of innovation in organizations. Account Manag Inf Technol 4(2):61-86

Coleman G, O'Connor R (2007) Using grounded theory to understand software process improvement: A study of Irish software product companies. Inf Softw Technol 49(6):654-667

Constantinides P, Barrett M (2006) Negotiating ICT development and use: The case of a telemedicine system in the healthcare region of Crete. Inf Organ 16(1):27-55

Darke P, Shanks G, Broadbent M (1998) Successfully completing case study research: Combining rigour, relevance and pragmatism. Inf Syst J 8(4):273-289

Dittrich Y, John M, Singer J, Tessem B (2007) Editorial for the special issue on qualitative software engineering research. Inf Softw Technol 49(6):531-539

Dittrich Y, Rönkkö K, Eriksson J, Hansson C, Lindeberg O (2008) Cooperative method development: Combining qualitative empirical research with method, technique and process improvement. Empir Softw Eng 13(3):231-260

Doolin B (1996) Alternative views of case research in information systems. Aust J Inf Syst 3(2):21-29

Easterbrook SM, Singer J, Storey M-A, Damian D (2008) Selecting empirical methods for software engineering research. In Shull F, Singer J, Sjøberg DIK (eds) Guide to advanced empirical software engineering. Springer, London, pp. 285-311

El Emam K, Koru AG (2008) A replicated survey of IT software project failures. IEEE Softw 25(5):84-90

Flyvbjerg B (2006) Five misunderstandings about case-study research. Qual Inquiry 12(2):219-245

Galliers RD, Swan JA (2000) There's more to information systems development than structured approaches: Information requirements analysis as a socially mediated process. Requir Eng 5(2):74-82

Gasson S (1999) A social action model of situated information systems design. Database Adv Inf Syst 30(2):82-97

Geertz C (1973) Thick description: Toward an interpretative theory of culture. In The interpretation of cultures: Selected essays. Basic Books, New York

Giddens A (1984) The constitution of society: Outline of the theory of structuration. University of California Press, Berkeley and Los Angeles

Glaser BG, Strauss A (1967) The discovery of grounded theory: Strategies for qualitative research. Aldine, Chicago

Glass RL, Vessey I, Ramesh V (2002) Research in software engineering: An analysis of the literature. Inf Softw Technol 44(8):491-506

Golden-Biddle K, Locke K (1993) Appealing work: An investigation of how ethnographic texts convince. Organ Sci 4(4):595-616

Guba EG (1990) The alternative paradigm dialog. In Guba EG (ed), The paradigm dialog. Sage, London, pp. 17-27

Harrison R, Badoo N, Barry E, Biffl S, Parra A, Winter B, Wuest J (1999) Directions and methodologies for empirical software engineering research. Empir Softw Eng 4(4):405-410

Höst M, Runeson P (2007) Checklists for software engineering case study research. In Proceedings of the First International Symposium on Empirical Software Engineering and Measurement (ESEM'07). IEEE Computer Society, Washington DC, pp. 479-481

Klein HK, Myers MD (1999) A set of principles for conducting and evaluating interpretive field studies in information systems. Manag Inf Syst Q 23(1):67-94

Langley A (1999) Strategies for theorizing from process data. Acad Manag Rev 24(4):691-710

Langley A, Truax J (1994) A process study of new technology adoption in smaller manufacturing firms. J Manag Stud 31(5):619-652

Lanzara GF (1999) Between transient constructs and persistent structures: Designing systems in action. J Strateg Inf Syst 8(4):331-349

Latour B (1987) Science in action: How to follow scientists and engineers through society. Harvard University Press, Cambridge, MA

Leonard-Barton D (1990) A dual methodology for case studies: Synergistic use of a longitudinal single site with replicated multiple sites. Organ Sci 1(3):248-266

Lethbridge TC, Singer J, Sim SE (2005) Studying software engineers: Data collection techniques for software field studies. Empir Softw Eng 10(3):311-341

Madsen S, Kautz K, Vidgen R (2006) A framework for understanding how a unique and local IS development method emerges in practice. Eur J Inf Syst 15(2):225-238

Markus ML, Lee AS (1999) Special issue on intensive research in information systems: Using qualitative, interpretive, and case methods to study information technology. Manag Inf Syst Q 23(1):35-38

Markus ML, Robey D (1988) Information technology and organizational change: Causal structure in theory and research. Manag Sci 34(5):583-598

McLeod L (2008) Understanding IS development and acquisition: A process approach. PhD thesis (http://hdl.handle.net/10292/644), Auckland University of Technology

McLeod L, MacDonell SG (in press) Factors that affect software systems development project outcomes: A survey of research. ACM Comput Surv

Miller J (2008) Triangulation as a basis for knowledge discovery in software engineering. Empir Softw Eng 13(2):223-228

Mingers J (2001) Combining IS research methods: Towards a pluralist methodology. Inf Syst Res 12(3):240-259

Myers MD, Newman M (2007) The qualitative interview in IS research: Examining the craft. Inf Organ 17(1):2-26

Nandhakumar J, Avison DE (1999) The fiction of methodical development: A field study of information systems development. Inf Technol People 12(2):176-191

Nandhakumar J, Jones M (1997) Too close for comfort? Distance and engagement in interpretive information systems research. Inf Syst J 7(2):109-131

Pentland BT (1999) Building process theory with narrative: From description to explanation. Acad Manag Rev 24(4):711-724

Perry DE, Porter AA, Votta LG (2000) Empirical studies of software engineering: A roadmap. In Proceedings of the Conference on the Future of Software Engineering. ACM, New York, NY, pp. 345-355

Perry DE, Sim SE, Easterbrook SM (2004) Case studies for software engineers. In Proceedings of the 26th International Conference on Software Engineering (ICSE'04). IEEE Computer Society, Washington DC, pp. 736-738

Perry DE, Sim SE, Easterbrook SM (2005) Case studies for software engineers. In Proceedings of the 29th Annual IEEE/NASA Software Engineering Workshop - Tutorial Notes. IEEE Computer Society, Los Alamitos CA, pp. 96-159

Pettigrew AM (1990) Longitudinal field research on change: Theory and practice. Organ Sci 1(3):267-292

Pettigrew AM (1997) What is processual analysis? Scandinavian Journal of Management 13(4):337-348

Robinson H, Segal J, Sharp H (2007) Ethnographically-informed empirical studies of software practice. Inf Softw Technol 49(6):540-551

Rönkkö K, Lindeberg O, Dittrich Y (2002) 'Bad practice' or 'bad methods': Are software engineering and ethnographic discourses incompatible? In Proceedings of the 2002 International Symposium on Empirical Software Engineering (ISESE'02). IEEE Computer Society, Washington DC, pp. 204-210

Royal Academy of Engineering (2004) The challenges of complex IT projects. Royal Academy of Engineering, London

Runeson P, Höst M (2009a) Guidelines for conducting and reporting case study research in software engineering. Empir Softw Eng 14(2):131-164

Runeson P, Höst M (2009b) Tutorial: Case studies in software engineering. In Bomarius F, Oivo M, Jaring P, Abrahamsson P (eds) Product-focused software process improvement (Proceedings of the 10th International Conference, PROFES 2009, Oulu, Finland, June 15-17). Springer, Berlin, pp. 441-442

Schultze U, Orlikowski WJ (2004) A practice perspective on technology-mediated network relations: The use of Internet-based self-serve technologies. Inf Syst Res 15(1):87-106

Seaman C (1999) Qualitative methods in empirical studies of software engineering. IEEE Trans Softw Eng 25(4):557-572

Segal J, Grinyer A, Sharp H (2005) The type of evidence produced by empirical software engineers. In Proceedings of the Workshop on Realising Evidence-Based Software Engineering (REBSE'05). ACM, New York, NY

Sieber JE (2001) Protecting research subjects, employees and researchers: Implications for software engineering. Empir Softw Eng 6(4):329-341

Sim SE, Singer J, Storey M-A (2001) Beg, borrow, or steal: Using multidisciplinary approaches in empirical software engineering research. Empir Softw Eng 6(1):85-93

Singer J, Vinson NG (2002) Ethical issues in empirical studies of software engineering. IEEE Trans Softw Eng 28(12):1171-1180

Sjøberg DIK, Dybå T, Jørgensen M (2007) The future of empirical methods in software engineering research. In Briand LC, Wolf AL (eds) Future of software engineering (FOSE'07). IEEE Computer Society, Los Alamitos, CA, pp. 358-378

Star SL (1989) The structure of ill-structured solutions: Boundary objects and heterogeneous distributed problem solving. In Gasser L, Huhns MN (eds) Distributed artificial intelligence (Vol. 2). Pitman Publishing, London, pp. 37-54

Subrahmanian E, Monarch I, Konda S, Granger H, Milliken R, Westerberg A, The N-Dim Group (2003) Boundary objects and prototypes at the interfaces of engineering design. Comput Support Co-op Work 12(2):185-203

Tolich M, Davidson C (1999) Starting fieldwork: An introduction to qualitative research in New Zealand. Oxford University Press, Auckland

Voss C, Tsikriktsis N, Frohlich M (2002) Case research in operations management. Int J Oper Prod Manag 22(2):195-219

Wainer J, Barsottini C (2007) Empirical research in CSCW - a review of the ACM/CSCW conferences from 1998 to 2004. J Braz Comput Soc 13(3):27-36

Wallace L, Keil M (2004) Software project risks and their effect on outcomes. Commun ACM 47(4):68-73

Walsham G (1993) Interpreting information systems in organizations. John Wiley and Sons, Chichester

Walsham G (1995) Interpretive case studies in IS research: Nature and method. Eur J Inf Syst 4(2):74-81

Walsham G (2006) Doing interpretive research. Eur J Inf Syst 15:320-330

Walsham G, Sahay S (1999) GIS for district-level administration in India: Problems and opportunities. Manag Inf Syst Q 23(1):39-66

Westrup C (1993) Information systems methodologies in use. J Inf Technol 8(4):267-275

Whittle A (2005) Preaching and practising 'flexibility': Implications for theories of subjectivity at work. Hum Relat 58(10):1301-1322

Wynekoop JL, Russo NL (1997) Studying system development methodologies: An examination of research methods. Inf Syst J 7(1):47-65

Yin RK (2003) Case study research: Design and methods, 3rd edn. Sage, Thousand Oaks, CA

Zuboff S (1988) In the age of the smart machine. Basic Books, New York

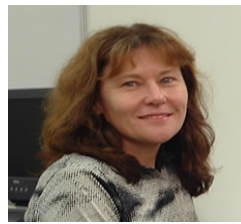

**Laurie McLeod** recently completed her PhD at Auckland University of Technology (AUT), New Zealand. Her doctoral research focused on a process approach for understanding software systems development and acquisition. Her research has been published in various journals, including *ACM Computing Surveys*, *Australasian Journal of Information Systems*, *Journal of Research and Practice in Information Technology* and *Journal of Systems and Information Technology*.

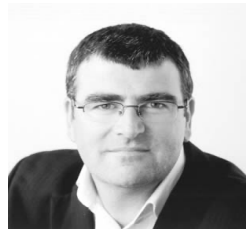

**Stephen G. MacDonell** is Professor of Software Engineering and Director of the Software Engineering Research Laboratory (SERL) at the Auckland University of Technology (AUT) in New Zealand. During his time at AUT Stephen has also held the roles of Head of the School of Information Technology and Associate Dean (Development). Stephen was awarded BCom(Hons) and MCom degrees from the University of Otago and a PhD from the University of Cambridge. He undertakes research in software metrics and measurement, project planning, estimation and management, software forensics, and the application of empirical analysis methods to software engineering data sets. He is a Member of the IEEE Computer Society and the ACM, and serves on the Editorial Board of *Information and Software Technology*.

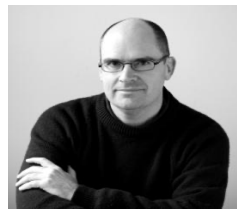

**Bill Doolin** is Professor of Technology and Organisation in the Business School at Auckland University of Technology (AUT), New Zealand. His research focuses on the processes that shape the adoption and use of information technologies in organizations. This has involved work on information systems in the public health sector and electronic business applications and strategies. His work has been published in international journals such as *Electronic Markets*, *Information Systems Journal*, *Journal of Global Information Management* and *Journal of Information Technology*.